\documentclass[graybox,vecphys,natbib]{svmult}

\usepackage{mathptmx}       
\usepackage{helvet}         
\usepackage{courier}        
\usepackage{type1cm}        
\usepackage{makeidx}         
\usepackage{graphicx}        
\usepackage{multicol}        
\usepackage[bottom]{footmisc}

\makeindex             

\usepackage{amsmath,amssymb}
\usepackage{url}

\newcommand{\pul}{\ensuremath{\frac{1}{2}}}

\newcommand{\pderiv}[2]{\frac{\partial #1}{\partial #2}}
\newcommand{\pderivl}[2]{{\partial #1}/{\partial #2}}
\newcommand{\deriv}[2]{\frac{{\text d}#1}{{\text d}#2}}

\newcommand{\zav}[1]{\left(#1\right)}

\newcommand{\melec}{\ensuremath{m_\text{e}}}
\newcommand{\velec}{\ensuremath{v_\text{e}}}

\newcommand{\Jiri}{Ji\v{r}\'{\i}}

\newcommand{\Kubat}{Kub\'at}

\begin{document}

\title*{Current status of NLTE analysis of stellar atmospheres}
\author{{\Jiri} {\Kubat}}
\institute{{\Jiri} {\Kubat} \at Astronomick\'y \'ustav AV \v{C}R,
Fri\v{c}ova 298, 251 65 Ond\v{r}ejov, Czech Republic,
\email{kubat@sunstel.asu.cas.cz}}

\maketitle

\abstract{Various available codes for NLTE modeling and analysis of hot
star spectra are reviewed.
Generalizations of standard equations of kinetic equilibrium and their
consequences are discussed.
}

\section{Importance of radiation}

Radiation is not only an information source about stellar atmospheres,
it has also the ability to interact with the matter in the atmosphere
and to alter its state.

Each photon has a momentum $h\nu/c$.
Consequently, when it is absorbed, the absorbing ion receives this
amount of momentum.
If it is scattered, the atom may receive from no momentum (if radiation
is scattered in the forward direction) to $2h\nu/c$, if radiation is
scattered backwards.
This momentum gain is redistributed to other particles by elastic
collisions and, as a consequence, acceleration of matter by radiation
occurs.

The energy of a photon ($h\nu$) can also be transferred to matter.
In the case of ionization, part of this energy is used to ionize the
atom and the rest goes to kinetic energy of electrons, $h\nu \rightarrow
h\nu_\text{ion}+\pul \melec \velec^2$.
In the case of excitation, all photon energy is used for this process, 
$h\nu \rightarrow h\nu_\text{exc}$.
In the case of the free-free absorption the photon energy causes
increase of the electron kinetic energy, $h\nu \rightarrow \pul \melec
\Delta\velec^2$.
Reverse processes (recombination, deexcitation, and free-free emission)
cause release of a photon and lowering the atomic internal or electron
kinetic energy (or both).
In the above mentioned cases of bound-free and free-free transitions
there is an energy exchange with electron kinetic energy.
In this case, subsequent elastic collisions re-establish the equilibrium
velocity distribution with a slightly different temperature.
Consequently, the thermal energy increases or decreases and we talk
about radiative heating or cooling.
If the energy exchange is with atomic internal energy, a change of
excitation or ionization balance occurs.
If the radiation changes of excitation and ionization states dominate,
then we have to take into account these effects explicitly and we have
to use the NLTE approximation and solve the kinetic equilibrium
equations.

\section{NLTE model atmosphere codes}

The solution of the kinetic equilibrium equations\footnote{These
equations are often referred to as the equations of statistical
equilibrium.
Here we use the term kinetic equilibrium equations as in the textbook
\citet{Hubeny:Mihalas:2014}.}
\citep[see][]{Hubeny:Mihalas:2014} for given radiative and collisional
rates is a relatively simple task, it is just a solution of set of
linear equations.
The task of NLTE modelling is more difficult.
It is the {\em simultaneous} solution of the equations of kinetic
equilibrium with other equations describing the stellar atmosphere.
These equations form a set of nonlinear integro-differential equations.

The basic task is to solve the equations of kinetic equilibrium together
with the radiative transfer equation to determine simultaneously the
level population numbers $n_i$ and the radiation field.
In this case, temperature $T(\vec{r})$, density $\rho(\vec{r})$, and
velocity $\vec{v}(\vec{r})$ are fixed.
This is the standard NLTE task, which may be applied also to trace
element NLTE calculations.

The task of calculation of a static NLTE model atmosphere is more
complicated.
It means adding two equations to the set of simultaneously solved
equations, namely the equation of hydrostatic equilibrium and the
equation of radiative equilibrium.
Consequently, temperature $T(r)$ and density $\rho(r)$ are not fixed,
but they are consistently calculated.
Then the results of solution of this set of nonlinear equations are the
NLTE level populations, radiation field, temperature, total density, and
electron density.
Adding the two structural equations causes significant slowing down the
convergence.

If we replace the hydrostatic equilibrium equation by the equation of
motion and add the continuity equation, we may solve hydrodynamic NLTE
model atmospheres and determine also the velocity field $v(r)$.
However, this full task is too complicated even for the case of
\mbox{1-D} atmospheres, so restricted problems are being usually solved.

Different codes solve different sets of equations using different
numerical approaches.
Below we list some representative examples of such codes.
Development of any NLTE model atmosphere code is a complicated and time
consuming task, which usually takes several years or even decades.
The description of such code is then usually spread over many
publications, if they exist at all.
In other cases, the most relevant information can appear only at a www
page, which may be even variable in time.
Therefore it is sometimes very difficult to pick up only a single clear
reference to a particular code, where everything important about the
code is described.
References in this paper were found with a belief that they give the
most appropriate reference to particular codes.
The author apologizes if he omitted references, which describe the code
better.
The url of the www page was added in the footnote for all cases when it
was known to the author.

A list of model atmosphere, NLTE, and radiative transfer codes was
compiled also by \cite{Hummer:Hubeny:1991} and \cite{sakhibullin:1996},
we refer the interested reader there.

\subsection{NLTE problem for a given structure}

There are many codes which solve the NLTE problem for a given (i.e.
fixed) atmospheric structure.
Here we list several of them, each of which represents possible method
of a solution of the problem.

One of the first codes which were able to solve the NLTE problem started
to be developed in mid-60s.
It was the code {\tt PANDORA}\footnote
{\url{http://www.cfa.harvard.edu/~avrett/pandora.html}}
\citep[see][]{Avrett:Loeser:2003}.
This code is based on the equivalent two-level atom approach and uses
\mbox{1-D} radiative transfer.
The code {\tt LINEAR} \citep{Auer:etal:1972} uses the complete
linearization method for solution of the multilevel line formation.
The code {\tt MULTI}\footnote {\url{http://folk.uio.no/matsc/mul22/}}
\citep{Carlsson:1986} uses the accelerated lambda iteration method for
solution of \mbox{1-D} NLTE multilevel problems both in static and
moving atmospheres in plane-parallel approximation.
\cite{Rybicki:Hummer:1991,Rybicki:Hummer:1992,Rybicki:Hummer:1994}
developed a method and a computer code {\tt MALI} for multilevel
radiative transfer using accelerated lambda iteration method.
The code described by \cite{Auer:etal:1994} and \cite{Fabiani:etal:1997}
solves \mbox{2-D} multilevel NLTE radiative transfer problem, using
again the accelerated lambda iteration method.

Various aspects of solution of the NLTE problem for trace elements
together with couple of codes were discussed in detail at the summer
school ``NLTE Line Formation for Trace Elements in Stellar Atmospheres''
\citep{Monier:etal:2010}.

\subsection{Static NLTE model atmospheres}

The pioneering work on the complete linearization method by
\cite{Auer:Mihalas:1969} led to a development of a computer code
for calculations of static plane parallel NLTE model atmospheres, which
was later described in \cite{Mihalas:etal:1975}.
An independent code using the same method was developed by
\cite{Kudritzki:1976}.
The most sophisticated NLTE static plane-parallel model atmosphere code
is the code TLUSTY\footnote
{\url{http://nova.astro.umd.edu/index.html}}, which was initially
developed using the complete linearization method
\citep{Hubeny:1975,Hubeny:1988}.
Now it combines the latter method with the accelerated lambda iteration
method \citep{Hubeny:Lanz:1995} and enables treatment of NLTE line
blanketing using the method of superlevels and superlines.
This code is accompanied with the code {\tt SYNSPEC}\footnote
{\url{http://nova.astro.umd.edu/Synspec49/synspec.html}}, which solves
the radiative transfer equation including precalculated NLTE populations
for given model atmosphere in detail.

Another code calculating NLTE line blanketed model atmosphere code is
the code {\tt PRO2} belonging to the T\"ubingen NLTE Model Atmosphere
Package \citep[{\tt TMAP}\rm
\footnote{\url{http://astro.uni-tuebingen.de/~TMAP/}},][]
{Werner:Dreizler:1999,Werner:etal:2003},
which focuses on modeling the atmospheres of high gravity stars (e.g.
white dwarfs) in NLTE.
We should also mention a multi purpose model atmosphere code {\tt
PHOENIX}\footnote
{\url{http://www.hs.uni-hamburg.de/EN/For/ThA/phoenix/}}$^{,}$\footnote
{\url{http://perso.ens-lyon.fr/france.allard/}}, which started as a
combination of two original codes, for cool stars \citep{Allard:1990},
and for novae atmospheres \citep{Hauschildt:1991}.

Static spherically symmetric model atmospheres were first calculated
using a complete linearization method  by \cite{Mihalas:Hummer:1974}.
An independent code using the same method was developed by
\cite{Gruschinske:1978}.
Another independent computer code {\tt ATA}\footnote
{\url{http://www.asu.cas.cz/~kubat/ATA/}} uses the accelerated lambda
iteration method and combines it with the linearization method to
calculate static NLTE model atmospheres of spherically symmetric stellar
atmospheres \citep{ATA1,ATA2,ATA3,ATA4,ATAsum}.

\subsection{NLTE wind model codes}

There exist several codes, which solve the NLTE model of the wind
assuming given density and velocity structure.
These codes solve simultaneously the equations of kinetic equilibrium
and the radiative transfer equation, which is mostly treated using  the
Sobolev approximation.
In some codes the more exact comoving frame solution of the radiative
transfer in spectral lines is used.
The continuum radiative transfer may be treated as in the static case
due to weak dependence of the opacity on frequency for continuum
transitions.

The code {\tt CMFGEN}\footnote
{\url{http://kookaburra.phyast.pitt.edu/hillier/web/CMFGEN.htm}}
\citep{Hillier:1987,Hillier:1990,Hillier:Miller:1998,Busche:Hillier:2005}
solves the equations of kinetic equilibrium and radiative transfer
equation in the comoving frame.
The velocity field is assumed to follow the $\beta$-law.
In addition, temperature structure of the models can be iteratively
determined.
This code may be considered as one of the top codes for solution of this
task.

The code {\tt PoWR}\footnote
{\url{http://www.astro.physik.uni-potsdam.de/~wrh/PoWR/}}
\citep{Hamann:1985,Hamann:Grafener:2004} solves also the equations of
kinetic equilibrium and radiative transfer equation in the comoving
frame.
The code was primarily designed for Wolf-Rayet stars atmosphere
modelling, however, nowadays it is also being applied to other types
stars with expanding atmospheres \citep[e.g.][]{modeling}.

Similarly the code {\tt FASTWIND}
\citep{Santolaya-Rey:1997,Puls:etal:2005}
solves the NLTE model for given velocity and density structure and has
been applied to many studies.
On the other hand, the code {\tt WM-basic}\footnote
{\url{http://www.usm.uni-muenchen.de/people/adi/adi.html}}
\citep[e.g.][]{Pauldrach:1987,Pauldrach:etal:1986,Pauldrach:etal:2012}
solves line blocked and blanketed NLTE model of stellar winds and also
their stationary hydrodynamic stratification.

The {\tt ISA-WIND} code
\citep{deKoter:etal:1993}
solves the equations of statistical equilibrium
and the radiative transfer equation in Sobolev approximation for given
velocity, density, and temperature structure.

There are also several codes focused on modelling supernovae, like
the code {\tt HYDRA} \citep{Hoflich:2003}, which combines radiative
hydrodynamics and NLTE radiative transfer.
NLTE radiative transfer in supernovae is solved using the Monte Carlo
method by \cite{Kromer:Sim:2009}.
The Monte Carlo method was used for NLTE calculations in circumstellar
disks in the code {\tt HDUST} by
\cite{Carciofi:Bjorkman:2006,Carciofi:Bjorkman:2008}.

\subsection{Hydrodynamic models with NLTE}

The most important task in stellar wind modelling is consistent
determination of the radiative force, which drives the wind.
Usually the so-called CAK approximation is being used, where the
radiative force is expressed with the help of three parameters $k$,
$\alpha$, and $\delta$ \citep{Castor:etal:1975,Abbott:1982}.
However, this approach is far from consistent even if the parameter $k$
is replaced by the more appropriate parameter $Q$ introduced by
\citet{Gayley:1995}.
There have been several attempts to improve calculation of the radiative
force using detailed list of lines.
First such calculations were done by \citet{Abbott:1982}.
Consistent calculation of the radiative force is enabled by the code
{\tt WM-basic} \citep[see][and references therein]
{Pauldrach:etal:2012}.
Monte Carlo calculations of the radiative force using NLTE radiative
transfer with the {\tt ISA-WIND} code were done by
\cite{Vink:etal:1999}.
Recently, \cite{nlte1,cmf1} developed a code for consistent solution of
wind hydrodynamic equations including equations of kinetic equilibrium.

\section{Generalized kinetic equilibrium}

Currently the most frequently applied mode of NLTE calculations is the
simultaneous solution of {\em static} (i.e. for macroscopic velocity
$\vec{v}=0$) and stationary ($\pderivl{}{t}=0$) equations of kinetic
equilibrium and the radiative transfer equation.
This basic set of equations may be extended by other constraint
equations for determination of temperature, density, and velocity.
However, there are many possible generalizations which go beyond this
``standard'' task.
More processes than simple one-electron excitation and ionization may be
included, and the assumption of static stationary medium in equilibrium
may be relaxed.

\subsection{Additional processes in the equations of kinetic
equilibrium}

A standard model atom consists of a number of energy levels, among which
collisional and radiative (both allowed and forbidden) transitions occur.
Usually, only transitions between levels of a particular ion and
ionization transitions from these levels to the ground level of the next
higher ion are considered.
This set of atomic levels and transitions is usually sufficient for
calculations of NLTE model atmospheres of hot stars.

However, there exist more processes which may change the state of the
atoms.
As an example we may consider Auger ionization, which occurs as a
consequence of a strong X-ray radiation.

Strong X-rays (or collisions) may expell an inner-shell electron.
This is followed by fluorescence or by another ionization.
The equations of kinetic equilibrium obtain additional term $n_i
\sum_{j>i} R_{ij}^\text{Auger}$, which describes this process,
\begin{equation}
- n_i \sum_{j\ne i} \zav{R_{ij} + C_{ij}} -
n_i \sum_{j>i} R_{ij}^\text{Auger} +
\sum_{j\ne i} n_j \zav{R_{ji} + C_{ji}}
=0
\end{equation}
The Auger ionization is considered as a two-electron ionization process.
States with inner-shell vacancy are not explicitly included in the
equations of kinetic equilibrium.
This form of the kinetic equilibrium equations was included to wind
ionization calculations by \cite{nlte3}.

\subsection{Full kinetic equilibrium equations}

The full form of the equations of kinetic equilibrium (in the classical
limit) reads \citep[][Eq.\,5-48]{Mihalas:1978},
\begin{displaymath}
{\pderiv{n_i}{t} + \nabla \cdot \zav{n_i \vec{v}}}
= \sum_{j\ne i} \zav{n_j P_{ji} - n_i P_{ij}}.
\end{displaymath}
As discussed in \cite{ja1}, the left hand side is usually neglected.
However, it may become important for dynamical atmospheres.
Non-equilibrium hydrogen ionization using the left hand side of the
equations of kinetic equilibrium was solved by
\citep{Leenaarts:etal:2007}.
The time derivative is important if relaxation timescale is longer than
dynamic timescales.
As examples may serve dynamic ionization of the solar chromosphere
\citep{Carlsson:Stein:2002} or rapidly expanding atmospheres of
supernovae \citep{Utrobin:Chugai:2005}.
This list of examples is far from complete and has to serve as an
indication of the importance of the full treatment of the equations of
kinetic equilibrium.

\subsection{Non-Maxwellian velocity distribution}

The presence of non-thermal electrons, which may happen, for example, as
a consequence of solar flares, is a reason why collisional transitions
do not support the system to evolve towards equilibrium distribution of
excitation and ionization states.
They rather cause that the populations differ from their equilibrium
values.
In this case, we may rewrite the equations of kinetic equilibrium as
\begin{equation}
- n_i \sum_{j\ne i} \zav{R_{ij} + C_{ij} + C_{ij}^\text{nt}}
+ \sum_{j\ne i} n_j \zav{R_{ji} + C_{ji}}
=0
\end{equation}
where the term $n_i C_{ij}^\text{nt}$ describes the non-thermal
collisional excitation.
This type of equations was used for inclusion of electron beams in solar
flares \citep{Kasparova:Heinzel:2002}.
To include the non-thermal collisional rates consistently, we have to
solve the kinetic equation for electrons instead of the assumption of
equilibrium distribution of their velocities.

\subsection{Polarization}

A significantly more general case is the case when we include polarized
radiation and polarization of atomic levels.
Then the radiation is described using the Stokes vector for intensities
\begin{equation}
\vec{I}=\zav{I, Q, U, V}^\text{T}.
\end{equation}
The radiative transfer equation becomes 
\begin{equation}
\deriv{\vec{I}}{s} = \vec{\eta} - \tens{\kappa} \vec{I},
\end{equation}
where the emission coefficient $\vec{\eta}$ is a vector quantity and the
opacity $\tens{\kappa}$ is a tensor quantity.

Instead of the equations of kinetic equilibrium we have to solve the
more general density matrix evolution equation,
\begin{equation}
\deriv{\tens{\rho}^K}{t}=0
\end{equation}
which, in addition to classical occupation numbers, solves also for
coherences between them.
For more details see \citet[and references
therein]{Stepan:TrujilloBueno:2013}.

\section{Summary}

The necessity to consider the equations of kinetic equilibrium (the NLTE
approach) is a consequence of radiation-matter interaction.
There exist many codes which can handle this relatively complicated
task.
Not all codes which claim ``NLTE'' solve the same task, one has always
to distinguish between codes for model atmosphere calculations and codes
for solution of the restricted NLTE problem (radiative transfer +
kinetic equilibrium).
The codes which calculate synthetic spectrum are much simpler.
They may enable input of NLTE population numbers, but their task is just
the formal solution of the radiative transfer equation (solution for
given opacity and emissivity).
There is also a number of codes which handle results of the previously
mentioned codes in a user-friendly manner.

Although NLTE model atmospheres and NLTE line formation problems offer
results, which are much closer to reality than those using pure LTE
approach, they are not the final step in generalization.
Much more has still to be done to include radiation-matter interactions
properly.

\begin{acknowledgement}
The author would like to thank Dr. Ewa Niemczura for inviting him to the
Spring School and he would also like to apologize her for the delay in
delivering manuscripts.
He is also grateful to both referees for their invaluable comments.
This work was partly supported by the project 13-10589S of the Grant
Agency of the Czech Republic (GA \v{C}R).
\end{acknowledgement}

\nocite{wrspec}
\newcommand{\aj}{Astron. J.}
\newcommand{\aap}{Astron. Astrophys.}
\newcommand{\aapr}{Astron. Astrophys. Rev.}
\newcommand{\apj}{Astrophys. J.}
\newcommand{\apjs}{Astrophys. J. Suppl. Ser.}
\newcommand{\araa}{Ann. Rev. Astron. Astrophys.}
\newcommand{\jqsrt}{J. Quant. Spectrosc. Radiat. Transfer}
\newcommand{\mnras}{Mon. Not. Roy. Astron. Soc.}

\bibliographystyle{spbasic}
\bibliography{kubat,wrspec,proc}

\end{document}